\documentclass[prl,twocolumn,showpacs]{revtex4}
\usepackage{graphicx}
\usepackage{amssymb}
\usepackage{amsmath}

\newcommand{\open}{\sphericalangle}

\begin{document}
\title{Dihadron fragmentation functions for large invariant mass}
\author{J.~Zhou and A.~Metz}
\affiliation{Department of Physics, Temple University,
 Philadelphia, Pennsylvania 19122-6082, USA}
\date{\today}
\begin{abstract}
Using perturbative Quantum Chromodynamics, we compute dihadron fragmentation 
functions for a large invariant mass of the dihadron pair.
The main focus is on the interference fragmentation function $H_1^{\open}$, 
which plays an important role in spin physics of the nucleon.
Our calculation also reveals that $H_1^{\open}$ and the Collins fragmentation
function have a closely related underlying dynamics.
By considering semi-inclusive deep-inelastic scattering, we further show that 
collinear factorization in terms of dihadron fragmentation functions, and 
collinear factorization in terms of single hadron fragmentation functions 
provide the same result in the region of intermediate invariant mass.
\end{abstract}
\maketitle

%
%
\noindent
{\it I. Introduction.}\,---\,Fragmentation functions (FFs) for quarks and gluons 
parameterize the hadronization taking place in high-energy scattering processes 
with identified strongly interacting particles in the final state.
The main focus is typically on FFs describing the transition of a parton into a 
single hadron --- see Ref.~\cite{Collins:1981uw} for a field-theoretic definition 
of such objects.
However, already in the late 1970's dihadron fragmentation functions (DiFFs) 
were introduced in order to quantify the hadron structure of 
jets~\cite{Konishi:1978yx}.
Moreover, it was shown that DiFFs are needed to obtain a consistent result for
the production of two hadrons in electron-positron annihilation when working
beyond leading order in perturbative Quantum Chromodynamics 
(QCD)~\cite{deFlorian:2003cg}.
In the meantime, DiFFs also play a considerable role in heavy ion physics --- 
see~\cite{Majumder:2004pt} and references therein.

In 1993, it was proposed~\cite{Collins:1993kq} that quark fragmentation into two 
hadrons can also be used to address the transversity distribution $h_1$ of the 
nucleon~\cite{Ralston:1979ys,Jaffe:1991kp}.
To this end, one can study the production of two hadrons in semi-inclusive
deep-inelastic scattering (DIS) in the current fragmentation 
region~\cite{Collins:1993kq}.
If the target is transversely polarized, there exists a correlation between the 
spin vector of the target and the orientation of the plane given by the momenta
of the two hadrons.
This observable contains the product of $h_1$ and a new fragmentation function
($H_1^{\open}$ in the notation of Ref.~\cite{Bianconi:1999cd}), which describes 
the strength of a correlation between the transverse polarization of the 
fragmenting quark and the orientation of the hadron plane.
Like the transversity, $H_1^{\open}$ is chiral-odd, and it results from the 
interference between two different production amplitudes why it is normally 
referred to as interference fragmentation function in the 
literature~\cite{Jaffe:1997hf}.

Data on the mentioned observable in semi-inclusive DIS have already been 
obtained by the HERMES and COMPASS 
Collaborations~\cite{Airapetian:2008sk,Wollny:2009eq}. 
The major difficulty is that, {\it a priori}, both $h_1$ and $H_1^{\open}$ 
are unknown. 
Existing models for $H_1^{\open}$~\cite{Jaffe:1997hf,Radici:2001na,Bacchetta:2006un} 
are still in a too early stage for getting a quantitative constraint on 
the transversity.
(For a related discussion we refer to~\cite{She:2007ht}.)
However, as was shown in~\cite{Artru:1995zu}, one can measure two back-to-back 
hadron pairs in electron-positron annihilation in order to get a handle on 
$H_1^{\open}$ --- see also~\cite{Boer:2003ya,Bacchetta:2008wb}, 
and Ref.~\cite{Vossen:2009xz} for preliminary data from the Belle Collaboration.
A combined analysis of the two processes allows one, in principle, to extract both 
unknown functions.
(First steps towards this goal are outlined in~\cite{Courtoy:2010qm}.)
Such a strategy would be very similar to the combined analysis of the Collins
effect~\cite{Collins:1992kk} in semi-inclusive DIS and in electron-positron 
annihilation~\cite{Efremov:2006qm,Anselmino:2007fs}, from which first information 
about the transversity was obtained~\cite{Anselmino:2007fs}.

The interference FF $H_1^{\open}$ and the Collins function ($H_1^{\perp}$ in
the notation of Refs.~\cite{Mulders:1995dh,Bacchetta:2006tn}) can be considered as 
complementary tools for getting a handle on the transversity distribution, with 
both having advantages and drawbacks.
An important advantage in the case of $H_1^{\open}$ is the fact that one can 
integrate over the total transverse momentum of the two hadrons in the final state,
leading to a collinear factorization formula.
In contrast, the Collins effect relies on factorization in terms of transverse 
momentum dependent parton correlators 
(TMD-factorization)~\cite{Collins:1981uk,Ji:2004wu,Collins:2004nx,Collins:2007ph},
which has additional technical complications.
On the other hand, when using $H_1^{\open}$, the dependence on the relative
transverse momentum of the two hadrons must be kept in order to have a well-defined
hadron plane.
This implies that $H_1^{\open}$ must also depend on the invariant mass $M_{hh}$ of 
the dihadron system.
If $M_{hh}$ is of the order of $\Lambda_{QCD}$, DiFFs are entirely non-perturbative 
objects.
In this kinematical region, one can only fit the DiFFs to experimental data or try
to estimate them by using some model for the strong interaction in the 
non-perturbative regime~\cite{Jaffe:1997hf,Radici:2001na,Bacchetta:2006un}.
(As a matter of principle, FFs cannot be computed in lattice gauge theory.)

In the present paper, we apply perturbative QCD in order to evaluate DiFFs for 
$M_{hh} \gg \Lambda_{QCD}$. 
The main focus is on the interference FF $H_1^{\open}$.
For large $M_{hh}$, the DiFFs can be expressed as a convolution of hard
coefficients and (collinear) single hadron fragmentation correlators.
In particular, the calculation determines the behavior of DiFFs as a function
of $M_{hh}$.
While the unpolarized DiFF $D_1$ drops like $1/M_{hh}^2$, the interference DiFF
$H_1^{\open}$ behaves like $1/M_{hh}^3$.
We also argue that $H_1^{\open}$ and the Collins function $H_1^{\perp}$ (at large
transverse momentum~\cite{Yuan:2009dw}) depend on the same two collinear twist-3 
fragmentation correlators, showing that the underlying dynamics of both functions 
is closely related.
In addition, we compute the (transverse) spin dependent cross section for dihadron 
production in semi-inclusive DIS for $\Lambda_{QCD} \ll M_{hh} \ll Q$
($Q$ denoting the virtuality of the exchanged photon) with single hadron FFs.
By comparing this result with the cross section obtained in the framework of DiFFs, 
we show that collinear factorization in terms of DiFFs holds as long as 
$M_{hh} \ll Q$.
In its spirit, our study is similar to recent work in which certain transverse 
momentum dependent parton correlators were evaluated for large transverse momenta, 
and the matching between collinear factorization and TMD-factorization was 
explicitly shown for intermediate transverse 
momenta~\cite{Ji:2006ub,Koike:2007dg,Yuan:2009dw} --- see also~\cite{Bacchetta:2008xw} 
for an overview.

%
\begin{figure}[t]
\includegraphics[width=6.5cm]{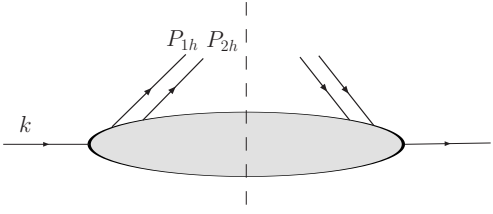}
\caption{Graphical representation of dihadron fragmentation function for a quark.}
\label{f:fig1}
\end{figure}
 
%
%
\noindent
{\it II. Kinematics and definition of dihadron fragmentation functions.}\,---\,We 
start by discussing the kinematics for the fragmentation of a quark into two hadrons
(displayed for the squared amplitude also in Fig.$\,$\ref{f:fig1}), 
\begin{eqnarray}
q(k) \rightarrow h_1(P_{1h}) + h_2(P_{2h}) + X \,.
\end{eqnarray}
We assume that the quark has a large light-cone minus momentum $k^-$.
For later convenience, we choose a reference frame in which one hadron has no 
transverse momentum.
In the hadron-1 frame, for instance, the light-cone components of the hadron 
momenta can be represented as
\begin{align}
P_{1h} & = \Big( 0, \, z_{1h} k^-, \, 0 \Big) \,,
\nonumber \\
P_{2h} & = \bigg( \frac{M_{hh}^2}{2 z_{1h} k^-}, \, z_{2h} k^-, \,  
                    \sqrt{\frac{z_{2h}} {z_{1h}}} M_{hh} \bigg) \,.
\end{align}
Neglecting the hadron masses, one readily verifies that 
$(P_{1h}+P_{2h})^2 = M_{hh}^2$.
We also introduce the total hadron momentum as well as the momentum difference
according to
\begin{equation}
P_{hh} = P_{1h} + P_{2h} \,, \quad
R = \frac{P_{1h}-P_{2h}}{2} \,.
\end{equation}
Their minus momenta are given by $P_{hh}^- = z_{hh} k^- , R^- = \hat{z}_{hh} k^-$,  
with $z_{hh} = z_{1h} + z_{2h}$ and $\hat{z}_{hh} = (z_{1h} - z_{2h})/2$.

The operator definition of the unpolarized DiFF and the interference DiFF, for a 
quark flavor $q$, reads
\begin{align}
& \int_{-\frac{1}{2}}^{\frac{1}{2}} \frac{d \hat z_{hh}}{32 (2 \pi)^3} 
  \frac{z_{hh}^2}{z_{1h}^2} \sum_X \int \frac{d y^+}{2\pi} e^{ik^- y^+} 
\nonumber \\
& \mbox{} \times \langle 0 | \psi^{q}(y^+) | P_{1h}, P_{2h}, X \rangle
  \langle P_{1h}, P_{2h}, X | \bar  \psi^{q}(0) | 0 \rangle
  \phantom{\frac{1}{1}}
\nonumber\\
& = \frac{\gamma^{+}}{2} D_1^q(z_{hh},M_{hh}^2)
  + \frac{\sigma^{\alpha +} R_{T \alpha}}{2 |\vec{R}_T|}
  H_1^{\open q}(z_{hh},M_{hh}^2) \,,
\end{align}
where a gauge link has been suppressed. 
Note that our definition of $H_1^{\open}$ differs from the one in 
Ref.~\cite{Bianconi:1999cd} by some prefactors.
We have integrated over the relative longitudinal momentum fraction $\hat{z}_{hh}$, 
which is needed if one wants to apply the collinear factorization discussed in the 
next section.

With these conventions, the parton model cross section for the production
of two hadrons in semi-inclusive DIS (with a transversely polarized proton),
$e p^{\uparrow} \to e h_1 h_2 X$, takes the form
\begin{align} \label{e:cs1}
& \frac{d \sigma}{dx_B dy d\phi_S dz_{hh} dM_{hh}^2 d\phi_R} =
  \frac{2 \alpha_{em}^2 s x_B}{Q^4} \sum_q e_q^2
\nonumber \\
& \mbox{} \times \bigg[ 
  \Big( 1 - y + \frac{y^2}{2} \Big) f_1^q(x_B) D_1^q(z_{hh},M_{hh}^2) 
  \nonumber\\ 
& \quad + (1-y) \sin(\phi_R+\phi_S) h_1^q(x_B) H_1^{\open q}(z_{hh},M_{hh}^2) \bigg] \,.
\end{align}
Equation~(\ref{e:cs1}) contains both the unpolarized cross section and one 
component depending on the transverse target polarization.
The azimuthal angle between $\vec{R}_T$ and the lepton plane is denoted by $\phi_R$, 
the azimuthal angle of the transverse spin vector of the proton is denoted by $\phi_S$,
while $x_B$ and $y$ are the commonly used DIS variables. 

\begin{figure}[t]
\includegraphics[width=8.5cm]{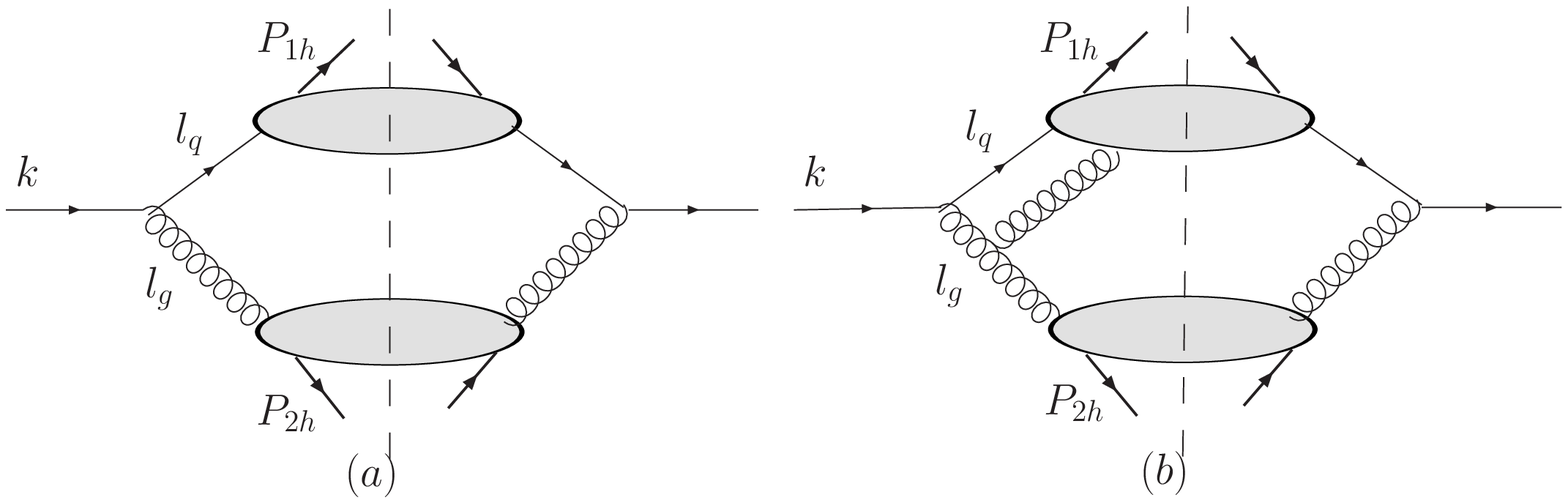}
\caption{Generic diagrams contributing to DiFFs for large invariant mass.
For the unpolarized DiFF $D_1$ the diagram~(a) needs to be considered, whereas 
in the case of $H_1^{\open}$ both diagrams are relevant.}
\label{f:fig2}
\end{figure}

%
%
\noindent
{\it III. Dihadron fragmentation functions at large invariant mass.}\,---\,When 
$M_{hh} \gg \Lambda_{QCD}$, dihadron fragmentation of a quark can be viewed as a 
two-step process: first, the quark splits into a quark (with momentum $l_q$) and 
a gluon (with momentum $l_g$), which is calculable in perturbative QCD.
Second, each of these two partons fragments into a single hadron.
This scenario is illustrated in Fig.$\,$\ref{f:fig2}.

We introduce momentum fractions $(z_1, z_2)$ through
\begin{equation}
l_q=\frac{P_{1h}}{z_1} \ ,\ l_g=\frac{P_{2h}}{z_2} \,,
\end{equation}
and define $\xi = z_{1h}/z_1$.
Evaluating the diagram in Fig.$\,$\ref{f:fig2}(a) one finds for the unpolarized
DiFF
\begin{align} \label{e:D1}
& D_1^q(z_{hh},M_{hh}^2) = \frac{\alpha_s}{2 \pi \, M_{hh}^2}  
  \int_{-\frac{1}{2}}^{\frac{1}{2}} d \hat z_{hh}
  \int_{z_{1h}}^{1-z_{2h}} \frac{d \xi}{\xi(1-\xi)} 
\nonumber \\
& \quad \mbox{}\times C_F \frac{1+\xi^2}{1-\xi} 
  D_1^{h_1/q}\Big(\frac{z_{1h}}{\xi}\Big) 
  D_1^{h_2/g}\Big(\frac{z_{2h}}{1-\xi}\Big) \,,
\end{align}
with $D_1^{h_1/q}$ and $D_1^{h_2/g}$ representing unpolarized single-hadron FFs. 
A second term, where hadron$\;$1 originates from the fragmentation of the gluon, 
is not written for brevity.
The result in Eq.~(\ref{e:D1}) shows that the DiFF $D_1$ drops like $1/M_{hh}^2$.
Note also that~(\ref{e:D1}) allows one to recover the inhomogeneous part of 
the evolution equation for $D_1$ by integrating over 
$M_{hh}^2$~\cite{Konishi:1978yx,deFlorian:2003cg}.

Computing $H_1^{\open}$ for large $M_{hh}$ is much more involved as one has to use 
collinear twist-3 factorization~\cite{Efremov:1981sh,Ellis:1982wd}.
One contribution arises from the diagram in Fig.$\,$\ref{f:fig2}(a), if the relative 
transverse momentum between the quark with momentum $l_q$ and the hadron$\;$1 is kept.
Also, 3-parton correlators including a transverse gluon field $A_{\perp}$ enter 
the calculation --- see the sample diagram in Fig.$\,$\ref{f:fig2}(b).
The correlators associated with these two contributions (the so-called 
$\partial_{\perp}$-contribution and the $A_{\perp}$-contribution) are of the 
form $\langle \bar{\psi} \partial_{\perp} \psi \rangle$ and
$\langle \bar{\psi} A_{\perp} \psi \rangle$, respectively.
In this paper, contributions from 3-gluon correlators are neglected since they do 
not affect any of our general results.
We leave this part, details of the calculation, and the discussion about a potential 
singularity in~(\ref{e:D1}) at $\xi = 0$ and $\xi = 1$ for future work~\cite{Zhou:prep}.
%

The calculation of the hard coefficient for the $A_{\perp}$-contribution is essentially
identical to the corresponding part in the treatment of the Collins function at large 
transverse momentum~\cite{Yuan:2009dw}.
On the other hand, to obtain the $\partial_{\perp}$-contribution is more complicated 
than for the Collins function evaluation.
To this end, we assign a (relative) transverse quark momentum according to
\begin{equation}
l_q = \frac{P_{1h}'}{z_1'} + l_{q\perp} \,,
\end{equation}
where $z_1' = z_1 + \delta z_1$, and $P_{1h}' = P_{1h} + \delta P_{1h}$. 
To keep $M_{hh}$ fixed, $P_{2h}$ must also change, i.e., 
$P_{2h}' = P_{2h} + \delta P_{2h}$.
In particular, $\delta P_{2h\perp} = - z_2 l_{q\perp}$. 
The kinematics is entirely determined by the constraints
\begin{align} \label{e:constr}
& P_{1h} \cdot \delta P_{1h} = 0 \,, \quad P_{2h} \cdot \delta P_{2h} = 0 \,,
\nonumber \\
& P_{1h} \cdot \delta P_{2h} + P_{2h} \cdot \delta P_{1h} = 0 \,,
\nonumber \\
& \delta P_{1h}^- + \delta P_{2h}^- = 0 \,, \quad \delta P_{1h\perp}=0 \,,
\end{align}
where we use the on-shell condition for the two hadrons, constraints from 
keeping $M_{hh}$ and $z_{hh}$ fixed, and the fact that we are working in the 
hadron-1 frame. 
(Recently, we used a related approach for computing a particular single spin 
asymmetry in the Drell-Yan process~\cite{Zhou:2010ui}.)
The solution to the set of equations in~(\ref{e:constr}) reads 
\begin{align}
\delta P_{1h} & =  \bigg( 0, \, 
  \frac{2 k^- l_{q\perp}}{P_{2h\perp}} \frac{z_2 z_{1h} z_{2h}}{z_{hh}}, \, 0 \bigg) \,,
\nonumber\\
\delta P_{2h} & =  \bigg( -\frac{P_{2h\perp} l_{q\perp}}{k^-} \frac{z_2}{z_{hh}}, \,
  -\frac{2 k^- l_{q\perp}}{P_{2h\perp}} \frac{z_2 z_{1h} z_{2h}}{z_{hh}}, \, 
  -z_2 l_{q\perp} \bigg) \,,
\nonumber\\
\delta z_1 & =  \frac{2}{z_{hh}}z_1 z_2 (z_2-z_1) (1-\xi) \frac{l_{q\perp}}{P_{2h\perp}} \,.
\end{align}
The partonic scattering amplitude $M$ depends now on $l_{q\perp}$.
We expand $M$,
\begin{align}
& M(P_{1h}',P_{2h}',z_1') =  M(P_{1h},P_{2h},z_1) 
  \phantom{\frac{1}{1}}
\nonumber \\
& \qquad
  + \frac{\partial M(P_{1h}',P_{2h}',z_1')}{\partial l_{q\perp}}
  \bigg|_{l_{q\perp}=0} l_{q\perp} + \ldots
\end{align}
and keep the term linear in $l_{q\perp}$ for obtaining the relevant (twist-3)
$\partial_{\perp}$-contribution~\cite{Ellis:1982wd,Zhou:prep}.

Both the $\partial_{\perp}$-contribution and the $A_{\perp}$-contribution can
be brought into a gauge invariant form.
After collecting all the pieces we find~\cite{Zhou:prep} 
\begin{align} \label{e:H1}
H_1^{\open q}(z_{hh},M_{hh}^2) & = \frac{\alpha_s}{2 \pi \, M_{hh}^3}  
  \int_{-\frac{1}{2}}^{\frac{1}{2}} \frac{d \hat z_{hh}}{\sqrt{z_{1h} z_{2h}}}
  \int_{z_{1h}}^{1-z_{2h}} \frac{d \xi}{\xi} 
\nonumber \\
& \mbox{} \times
  A^{h_1/q}(\hat z_{hh},\xi)
  D_1^{h_2/g}\Big(\frac{z_{2h}}{1-\xi}\Big) \,,
\end{align}
where the function $A$ is defined as
\begin{align}
A & = C_F \bigg[ \bigg( z_1^3 \frac{\partial}{\partial z_1} \frac{\hat H(z_1)}{z_1^2} \bigg)
  2 \xi^2 \frac{z_1-z_2}{z_{hh}}  
  + \hat H(z_1) \frac{2 \xi^2}{1-\xi} \bigg]
\nonumber \\
& + \int \frac{d\bar{z}_1}{\bar{z}_1^2} 
  PV \bigg(\frac{1}{\frac{1}{z_1}-\frac{1}{\bar{z}_1}} \bigg) 
  \hat H_F(z_1,\bar{z}_1)
\nonumber \\
& \qquad \mbox{} \times 
  \bigg[ -C_F \frac{2z_{1h}}{z_1} \bigg( 1 + \frac{z_{1h}}{\bar{z}_1} 
  - \frac{z_{1h}}{z_1} \bigg)
\nonumber \\
& \qquad - \frac{C_A}{2}\frac{2z_{1h}}{z_1}
  \frac{z_1 \bar{z}_1 (z_1 + \bar{z}_1) - z_{1h} (z_1^2 + \bar{z}_1^2)}
  {z_1(z_1 - \bar{z}_1)(\bar{z}_1 - z_{1h})} \bigg] \,.
\end{align}
This result shows that $H_1^{\open}$ behaves like $1/M_{hh}^3$ for large $M_{hh}$.
It also reveals an intimate connection between $H_1^{\open}$ and the Collins 
function (at large transverse momentum)~\cite{Yuan:2009dw}, as both functions 
depend on the same collinear twist-3 correlation functions $\hat{H}$ and 
$\hat{H}_F$, which we take in the definition of Ref.~\cite{Yuan:2009dw}.
The hard coefficients (of the $\partial_{\perp}$-contribution) differ in both
cases.

%
%
\noindent
{\it IV. Single spin asymmetry for dihadron production in semi-inclusive DIS}\,---\,In 
this section, we investigate the validity of the factorization formula~(5) for the 
cross section of dihadron production in semi-inclusive DIS for the case
$\Lambda_{QCD} \ll M_{hh} \ll Q$.
Though one might expect this factorization to hold, to the best of our knowledge
no explicit supportive calculation exists.
We focus on the discussion of the spin-dependent component of the cross section,
which is nontrivial already at lowest order in perturbation theory.
%
\begin{figure}[t]
\includegraphics[width=8.5cm]{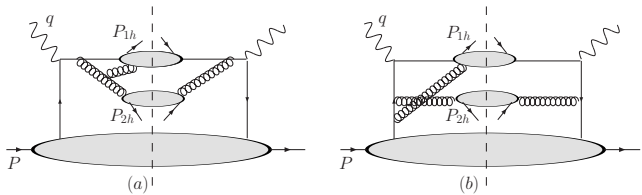}
\caption{Sample diagrams contributing to dihadron production in semi-inclusive 
DIS. Diagram~(a) generates a leading power contribution and diagram~(b), in the
light-cone gauge, is power-suppressed for $M_{hh} \ll Q$.}
\label{f:fig3}
\end{figure}
%

For $M_{hh}$ of the order of $Q$, one can use collinear factorization in terms 
of single hadron fragmentation correlators.
Sample diagrams are shown in Fig.$\,$3.
We have evaluated the cross section for this kinematics, and then expanded the 
result for $\Lambda_{QCD} \ll M_{hh} \ll Q$.
For factorization in terms of DiFFs to hold, the result has to match with (5), 
if $H_1^{\open}$ for $M_{hh} \gg \Lambda_{QCD}$ from~(\ref{e:H1}) is inserted.
By inspecting the Feynman diagrams it becomes obvious that this matching is
indeed nontrivial.
For instance, diagram~(b) in Fig.$\,$3 has no counterpart in the calculation of 
$H_1^{\open}$ for large $M_{hh}$.
However, it turns out that for $M_{hh} \ll Q$ this diagram is power-suppressed.
This is true only in the light-cone gauge $A^{-} = 0$ which we use for this 
analysis.
In a covariant gauge, the treatment gets more involved~\cite{Zhou:prep}. 
In the end we indeed find a matching, showing the consistency of the factorization
in terms of DiFFs for $M_{hh} \gg \Lambda_{QCD}$ as long as $M_{hh} \ll Q$.
For the unpolarized cross section the corresponding analysis is trivial to lowest 
order, but also becomes nontrivial ones loop corrections are included.

%
%
{\it IV. Conclusions.}\,---\,
In this paper, we have studied DiFFs for a large invariant mass $M_{hh}$ of the
dihadron pair, where perturbative QCD can be applied.
The main focus has been on the interference DiFF $H_1^{\open}$, which drops like
$1/M_{hh}^3$ for large $M_{hh}$ and is related to the same universal twist-3
collinear fragmentation correlators that describe the Collins FF $H_1^{\perp}$ (at
large transverse momentum).
The analysis also predicts that the transverse single spin asymmetry for dihadron
production in semi-inclusive DIS behaves like $1/M_{hh}$. 
The preliminary COMPASS data~\cite{Wollny:2009eq}, ranging up to 
$M_{hh} \approx 2\, GeV$, are in agreement with this general result.
We expect a corresponding behavior for the so-called Artru-Collins asymmetry
in electron-positron annihilation~\cite{Artru:1995zu}, for which preliminary data 
from Belle exist~\cite{Vossen:2009xz}.
For the case of semi-inclusive DIS we have also shown explicitly, to lowest nontrivial 
order in perturbation theory, that collinear factorization in terms of DiFFs is 
consistent for large $M_{hh}$ provided that $M_{hh} \ll Q$.

This work is supported by the NSF under Grant No.~PHY-0855501.

%
%

\end{document}